\documentclass[aps,prl,reprint,twocolumn,amsmath,amssymb,citeautoscript]{revtex4-1}
\usepackage{graphicx}
\usepackage{dcolumn}
\usepackage{bm}
\usepackage{latexsym,epsfig}
\usepackage{graphicx}
\usepackage{verbatim}
\usepackage{comment}
\usepackage{amsmath}
\usepackage{amssymb}
\usepackage{physics}
\usepackage{stmaryrd}
\usepackage{color}
\usepackage{epstopdf}
\usepackage{grffile}
\usepackage{lipsum}
\usepackage{enumitem}
\usepackage[normalem]{ulem}
\usepackage{subfigure}

\DeclareGraphicsExtensions{.eps}

\newcommand{\beq}{\begin{equation}}
	\newcommand{\eeq}{\end{equation}}
\newcommand{\bea}{\begin{eqnarray}}
	\newcommand{\eea}{\end{eqnarray}}
\newcommand{\ben}{\begin{eqnarray*}}
	\newcommand{\een}{\end{eqnarray*}}
\newcommand{\bfig}{\begin{figure}}
	\newcommand{\efig}{\end{figure}}

\usepackage{hyperref}
\hypersetup{
	colorlinks=true,      
	urlcolor=blue,
	citecolor=blue,
	linkcolor=blue
}

\begin{document}
	
	\title{Interplay of Electron–Phonon Coupling, Dissipative Phonon Bath, and Electron–Electron Interaction in a Triangular Quantum-Dot Trimer}
	
	\author{Hemant Kumar Sharma}
	\affiliation{School of Physical Sciences, National Institute of Science Education and Research, Jatni 752050, India}
	\date{\today}
	
	\begin{abstract}
	Nonequilibrium charge transport through a trimer molecular transistor (TMT) composed of three quantum dots arranged in a triangular geometry, which is placed on a substrate, has been studied in the presence of electron-electron and electron-phonon interactions. The entire system is described by an extended Anderson–Holstein–Caldeira–Leggett Hamiltonian, in which the Caldeira–Leggett term accounts for phonon damping arising from the coupling between the molecular vibrations and the substrate phonon bath. The electron–phonon interaction is treated nonperturbatively using the Lang–Firsov canonical transformation, while the electron–electron interaction is incorporated at the mean-field level.  Keldysh nonequilibrium Green’s function (NEGF) framework is used to study the transport properties, allowing us to calculate the spectral function, tunneling current, and differential conductance of the trimer molecular transistor. The formalism enables systematic evaluation of the effects of Coulomb interaction, electron–phonon coupling, and dissipation on the device’s electronic transport characteristics.
	\end{abstract}
	
	\maketitle
	
	\section{Introduction}
Semiconductor quantum dots (QDs) provide a universal platform for analyzing quantum transport, strong correlations, and vibronic effects at the nanoscale~\cite{Kouwenhoven1997,Hanson2007,Datta1995}. Improvements in nanofabrication have made it easier to make coupled-dot architectures like double and triple quantum dots, where electron coherence, interference, and Coulomb interactions create a lot of interesting transport phenomena. ~\cite{Fujisawa2006,Gaudreau2009,vanDerWiel2002}.

Among these, triangular triple quantum-dot or trimer molecular configurations have attracted particular attention due to their inherent geometric frustration and multiple tunneling pathways~\cite{Amaha2012,Seo2013,Takakura2014,Trocha2010}. The closed-loop geometry allows for direct and indirect tunneling channels, leading to quantum interference and symmetry-protected degeneracies that strongly influence conductance and level hybridization~\cite{Numata2009,Oguri2011}.

When such nanoscale systems are embedded on or coupled to insulating substrates, the interaction between electrons and local vibrational modes becomes unavoidable. The electron–phonon (e–ph) coupling locally modulates the dot potential and gives rise to phonon-assisted tunneling, vibronic sidebands, and polaron formation~\cite{Galperin2007,LangFirsov1963,Hohenadler2007,Koch2005}. The canonical Lang–Firsov transformation (LFT) provides a nonperturbative route to account for these effects, yielding renormalized onsite energies and tunneling amplitudes that depend exponentially on the e–ph coupling strength~\cite{LangFirsov1963,Mahan2000}. Experimental evidence of such polaronic phenomena has been observed in suspended carbon-nanotube QDs and molecular junctions~\cite{Hohberger2003,Koch2005,Mitra2004}, where phonon sidebands and phonon blockade appear in current and conductance spectra.

In realistic environments, however, the phonon modes of QDs are not isolated. They couple to substrate or surrounding lattice vibrations, which act as a dissipative phonon reservoir. The Caldeira–Leggett model~\cite{Caldeira1983,Leggett1987,Weiss2012} captures this coupling by introducing a spectral density that governs frequency renormalization and damping of the local phonon mode. The resulting phonon softening and energy relaxation profoundly modify charge transport by broadening sidebands and suppressing coherence~\cite{Leijnse2010,Braig2003,Sharma2024SciRep}. The competition between electron–phonon interaction and bath-induced damping thus determines whether transport is coherent, phonon-assisted, or dissipative.

Electron–electron (e–e) interactions introduce an additional layer of complexity. The onsite Coulomb repulsion $U$ leads to Coulomb blockade and spin or charge correlations that can suppress or shift resonant transport windows~\cite{Hanson2007,Meir1991}. In multi-dot geometries, mean-field treatments of $U$ yield effective level shifts that modify interdot coherence and current–voltage response~\cite{Bruder1996,Cornaglia2004,Trocha2010}. The simultaneous inclusion of e–ph coupling, dissipation, and Coulomb interaction in a triangular geometry, however, has not been extensively explored.

In this work, we investigate charge transport through a trimer molecular transistor (TMT)—a triangular triple quantum-dot system coupled to a substrate phonon bath—by treating electron–phonon coupling, phonon damping, and on-site Coulomb interaction within a unified nonequilibrium Green’s function framework. The electron–phonon interaction is handled nonperturbatively using the Lang–Firsov transformation, while the substrate is modeled via a Caldeira–Leggett bath with damping parameter $\gamma$ that renormalizes the effective phonon frequency $\tilde{\omega}_0$. The on-site interaction $U$ is treated within the mean-field approximation, appropriate for the weak to intermediate correlation regime. This formulation allows us to compute spectral functions, steady-state currents, differential conductance maps, and dot occupations self-consistently under finite bias.

Our results reveal how the interplay between electron-phonon, damping $\gamma$ , and electron-electron interaction $U$  strength controls electronic transport. Increasing electron-phonon interaction strength $\lambda$ enhances phonon sidebands and induces polaronic suppression of current, while larger $\gamma$ leads to damping and eventual localization. Finite $U$ introduces Coulomb-split sidebands and nonlinear current–voltage characteristics with regions of negative differential resistance. The combined analysis of the spectral function, occupation, and two-dimensional conductance maps provides a complete microscopic picture of correlated, dissipative, and vibronic transport in trimer quantum-dot systems. These findings are directly relevant for molecular transistors, nanoscale interferometers, and phonon-assisted tunneling devices.
	
	\section{Hamiltonian Description}
\begin{figure}[t]
	\centering
	\includegraphics[width=1\columnwidth]{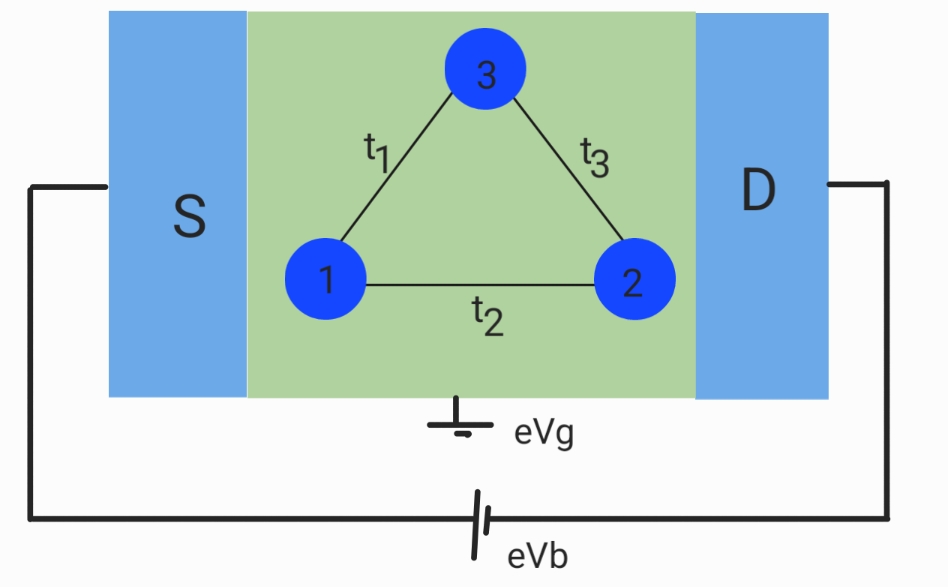}
	\caption{Schematic representation of the trimer molecular transistor (TMT). 
		The central region consists of three quantum dots (QDs) forming a triangular configuration, 
		coupled to the source (S) and drain (D) electrodes. 
		The inter-dot tunneling amplitudes are denoted by $t_1$, $t_2$, and $t_3$, while $eV_g$ and $eV_b$ represent the gate and bias voltages, respectively.}
	\label{fig:Trimer}
\end{figure}

In Fig.~\ref{fig:Trimer}, we present the schematic diagram of the trimer molecular transistor (TMT) device under consideration. 
The central region of this device consists of three quantum dots (QDs) arranged in a triangular geometry. 
QD$_1$ is connected to the source (S), QD$_2$ to the drain (D), while QD$_3$ is coupled to both QD$_1$ and QD$_2$. 
All QDs are placed on an insulating substrate that hosts a collection of uncoupled simple harmonic oscillators 
and acts as a phonon bath. Each QD possesses a single lattice vibrational mode that interacts with its electronic degree of freedom 
through a Holstein-type electron--phonon (e--p) coupling. 
The phonons of the QDs are further linearly coupled to the phonons of the substrate. 
This coupling is modeled using the Caldeira--Leggett (CL) Hamiltonian, 
which introduces dissipation into the tunneling current through the device. 
We consider a three-dot system with a single energy level $\varepsilon$ per dot, 
coupled via inter-dot hopping amplitudes $t_1$, $t_2$, and $t_3$. 
QD$_1$ (QD$_3$) is connected to the left (right) metallic lead. 
The total Hamiltonian of the system is given by:

	\begin{equation}
		H = H_{\text{leads}} + H_{\text{QD}} + H_B + H_{\text{tun}},
	\end{equation}
	where $H_{\text{leads}} = \sum_{\alpha=L,R}\sum_{k\sigma}\epsilon_{k\alpha} c_{k\alpha\sigma}^\dag c_{k\alpha\sigma}$ describes noninteracting leads with Fermi functions $f_{\alpha}(\omega)$ and chemical potentials $\mu_\alpha$, and
	\begin{align}
		H_{\text{tun}} = \sum_{k\sigma} \big(
		V_{kL} c_{kL\sigma}^\dag d_{1\sigma} + 
		V_{kR} c_{kR\sigma}^\dag d_{3\sigma} + \text{H.c.} \big),
	\end{align}
	defines the lead–dot tunneling with hybridization functions 
	$\Gamma_{\alpha}(\omega) = 2\pi \sum_k |V_{k\alpha}|^2 \delta(\omega-\epsilon_{k\alpha})$~\cite{Meir1991,Datta1995}.  
	The molecular Hamiltonian includes on-site energy, Coulomb repulsion, and electron–phonon (el–ph) coupling:
	\begin{align}
		H_{\text{QD}} &= 
		\sum_{i,\sigma} \varepsilon\, d_{i\sigma}^\dag d_{i\sigma}
		+ U \sum_i n_{i\uparrow} n_{i\downarrow}
		+ \sum_{\langle i,j\rangle,\sigma} t_{ij} d_{i\sigma}^\dag d_{j\sigma} \notag\\
		&+ \sum_i \left(\frac{p_i^2}{2m_0} + \frac{1}{2}m_0\omega_0^2 x_i^2 \right)
		+ \sum_{i,\sigma} g_i n_{i\sigma} x_i,
	\end{align}
	where $g_i$ denotes the local el–ph coupling strength~\cite{LangFirsov1963,Galperin2007}.  
	The substrate is modeled by a Caldeira–Leggett bath,
	\begin{align}
		H_B = \sum_j \!\left(\frac{p_j^2}{2m_j} + \frac{1}{2}m_j\omega_j^2 x_j^2\right) 
		+ \sum_{i,j} \beta_j x_i x_j,
	\end{align}
	whose spectral density $J(\omega)=\sum_j \frac{\beta_j^2}{2m_j\omega_j^2}\delta(\omega-\omega_j)$ renormalizes the phonon frequency as 
	$\tilde{\omega}_0^2=\omega_0^2-(2/m_0)\!\int\! d\omega J(\omega)/\omega$~\cite{Caldeira1983,Sharma2024SciRep}.  
	For a Lorentz–Drude form $J(\omega)=2m_0\gamma\omega/[1+(\omega/\omega_c)^2]$, 
	$\tilde{\omega}_0^2=\omega_0^2-2\pi\gamma\omega_c$~\cite{Sharma2024SciRep,Weiss2012}.
	
	The el–ph coupling is treated nonperturbatively using the Lang–Firsov unitary transformation 
	$U=\exp[-\sum_i (g_i/\omega_0)n_i(b_i^\dag-b_i)]$,  
	which shifts phonon coordinates and renormalizes electronic operators as 
	$U^\dag d_i U = d_i X_i$, $X_i = \exp[(g_i/\omega_0)(b_i^\dag-b_i)]$~\cite{LangFirsov1963,Hohenadler2007}.  
	The hopping term transforms as 
	$t_{ij} d_{i\sigma}^\dag d_{j\sigma}\!\to\! t_{ij} d_{i\sigma}^\dag X_i^\dag X_j d_{j\sigma}$,  
	and at $T=0$, $\langle X_i^\dag X_j\rangle=e^{-(g/\tilde{\omega}_0)^2}$ gives the renormalized hopping 
	$t_{\text{eff}} = t\, e^{-(g/\tilde{\omega}_0)^2}$~\cite{Koch2005,Sharma2024SciRep}.  
	The transformed Hamiltonian becomes
	\begin{align}
		\widetilde{H} = H_{\text{leads}} + \widetilde{H}_{\text{QD}} + H_{\text{tun}},
	\end{align}
	with 
	$\widetilde{H}_{\text{QD}}=\sum_{i\sigma}\tilde{\epsilon}_i d_{i\sigma}^\dag d_{i\sigma}
	+\sum_{\langle i,j\rangle\sigma}\tilde{t} d_{i\sigma}^\dag d_{j\sigma}
	+\tilde{U}\sum_i n_{i\uparrow}n_{i\downarrow}$,
	where $\tilde{\epsilon}_i=\epsilon_i-\hbar\omega_0(g_i/\omega_0)^2$, 
	$\tilde{U}=U-\hbar\omega_0(g_i/\omega_0)^2$, and $\tilde{t}=t_{\text{eff}}$~\cite{LangFirsov1963,Jeon2003}.
	
	The tunneling current from the source lead is obtained from 
	$J_S = -(ie/\hbar)\langle[N_S,H]\rangle$,  
	where $N_S=\sum_{k\sigma} c_{kL\sigma}^\dag c_{kL\sigma}$.  
	Using the Keldysh formalism and Langreth continuation, one obtains
	\begin{align}
		J_{S\sigma} &= \frac{ie}{2\pi\hbar}\!\int\! \Gamma_S(\epsilon)
		\big[ G_{QD}^{<}(\epsilon)
		+ f_S(\epsilon)(G_{QD}^{r}-G_{QD}^{a}) \big] d\epsilon,
	\end{align}
	and an analogous expression for $J_{D\sigma}$ from the drain~\cite{Keldysh1965,Haug1996,Meir1991}.  
	At steady state $J=J_S=-J_D$, leading to
	\begin{align}
		J = \frac{e}{\hbar} \!\int\! d\epsilon\,[f_S(\epsilon)-f_D(\epsilon)] 
		T(\epsilon),
	\end{align}
	where $T(\epsilon)=\Gamma_S(\epsilon)\Gamma_D(\epsilon)|G_{QD}^r(\epsilon)|^2$ 
	is the transmission function~\cite{Datta1995,Meir1992}.  
	For symmetric coupling $\Gamma_S=\Gamma_D=\Gamma$ and wide-band limit,  
	$J=(e\Gamma/\hbar)\!\int\! d\epsilon [f_S(\epsilon)-f_D(\epsilon)] A(\epsilon)$,  
	where $A(\epsilon)=i(G_{QD}^r-G_{QD}^a)$ is the spectral function.  
	
	The retarded, lesser, and greater Green functions for dot 1 include phonon dressing via
	\begin{align}
		G_{QD}^r(t-t') &= -i\theta(t-t') 
		\langle \{ d_1(t), d_1^\dagger(t') \} \rangle 
		\langle X_1(t) X_1^\dagger(t') \rangle_{ph},
	\end{align}
	where 
	$\langle X_1^\dagger(t)X_1(t')\rangle_{ph}
	= \sum_{l=-\infty}^{\infty} L_l e^{il\omega_0(t-t')}$~\cite{Mahan2000,Sharma2024SciRep}, 
	and $L_l = e^{-g^2(1+2N_{ph})}\sum_{l'} I_{l'}(z')e^{l'\beta\hbar\omega_0/2}$ 
	with $z'=2g^2\sqrt{N_{ph}(N_{ph}+1)}$~\cite{Mahan2000,Chen2005}.  
	At $T=0$, $L_l=e^{-g^2} g^{2l}/l!$ $(l\ge0)$~\cite{Mahan2000}.  
	The phonon-assisted spectral function follows as
	\begin{align}
		A(\varepsilon)=\sum_{l=-\infty}^{\infty} iL_l
		[\widetilde{G}_{QD}^r(\varepsilon-l\hbar\omega_0)
		-\widetilde{G}_{QD}^a(\varepsilon+l\hbar\omega_0)],
	\end{align}
	and $\widetilde{G}_{QD}^<(\varepsilon)=
	\widetilde{G}_{QD}^r(\varepsilon)\Sigma^<(\varepsilon)\widetilde{G}_{QD}^a(\varepsilon)$
	with $\Sigma^<_\alpha(\varepsilon)=if_\alpha(\varepsilon)\Gamma_\alpha(\varepsilon)$~\cite{Haug1996,Datta1995}.
	This formalism captures both polaronic renormalization and phonon sidebands in the tunneling spectra of the TQD system~\cite{Koch2005,Galperin2007,Sharma2024SciRep}.
	
	After calculating the self-energy, we calculate the retarded and advanced Green functions 
	$G_{QD}^{r(a)}$ using the equation of motion technique:
	\begin{equation}
		\begin{aligned}
			G^r_{QD}(E) &=
			\frac{(E-\varepsilon)\left(E-\varepsilon+\tfrac{i}{2}\Gamma_R\right) - t_3^2}{
				\left(E-\varepsilon+\tfrac{i}{2}\Gamma_L\right)
				\left[(E-\varepsilon)\left(E-\varepsilon+\tfrac{i}{2}\Gamma_R\right) - t_3^2\right]} \\[4pt]
			&\quad - 
			\frac{t_2^2\left(E-\varepsilon+\tfrac{i}{2}\Gamma_R\right)
				+ 2t_1t_2t_3 + t_1^2(E-\varepsilon)}{
				\left(E-\varepsilon+\tfrac{i}{2}\Gamma_L\right)
				\left[(E-\varepsilon)\left(E-\varepsilon+\tfrac{i}{2}\Gamma_R\right) - t_3^2\right]}.
		\end{aligned}
	\end{equation}
	\section{Result}
	We have assumed that QD$_{1,2}$ is symmetrically connected to both the source and the drain. 
	The energy level of each quantum dot (QD) is taken to be zero, and all energies are measured in units of the phonon energy $\hbar \omega_0$. 
	For most of our calculations, we consider $\Gamma = 0.2$, $eV_g = 0$, $eV_m = 0.5$, $eV_b = 0.5$,$t_1=t_2 = t_3 = 0.3$ and $U = 1$. 
	The Coulomb interaction is treated within the mean-field Hartree--Fock approximation. 
	Because of the electron--phonon (e--p) interaction, the onsite Coulomb interaction energy $U$ is substantially reduced due to the polaronic effect, 
	and thus the use of the mean-field approximation is justified for moderately large Coulomb interactions. 
	Hence, our results lie well outside the Kondo regime. 
	We also assume that the electron charge density in both the source and the drain remains constant.
	
\begin{figure}[t]
		\centering
		\includegraphics[width=1\columnwidth]{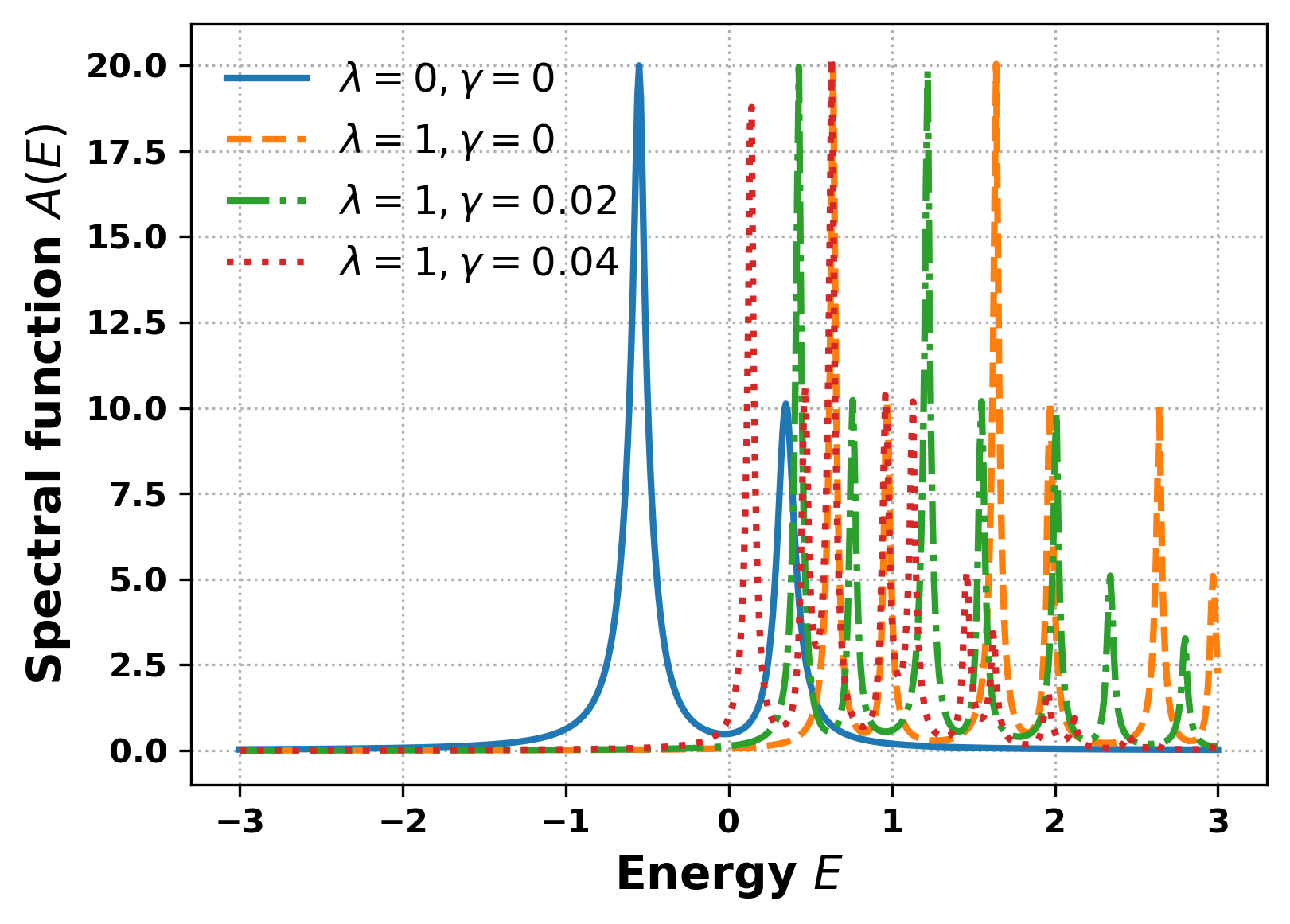}
		\caption{Spectral function $A(E)$ as a function of energy $E$ for different values of electron--phonon coupling strength $\lambda$ and dissipation parameter $\gamma$. }
		\label{fig:SpectralFunction}
	\end{figure}

	Fig~\ref{fig:SpectralFunction} displays the behaviour of the spectral function $A(E)$ as a function of energy $E$ for different values of electron--phonon coupling strength $\lambda$ and dissipation parameter $\gamma$. 
	The blue solid line ($\lambda = 0, \gamma = 0$) corresponds to the non-interacting case, showing two sharp peaks that represent the discrete electronic levels of the isolated system. 
	When $\lambda = 1$ and $\gamma = 0$ (orange dashed line), phonon sidebands appear, resulting from electron--phonon interactions that split the spectral weight into multiple peaks. 
	Increasing $\gamma$ (green dot-dashed and red dotted lines) introduces dissipation due to coupling with the phonon bath, leading to a broadening and reduction in intensity of these peaks. Thus, $\lambda$ creates multiple phonon-assisted channels, while $\gamma$ induces relaxation and dephasing, both reshaping the molecular density of states.
\begin{figure}[t]
	\centering
	
	\subfigure[Current $J(eV_b)$ and differential conductance $G = dJ/d(eV_b)$ for $U = 0$.]{
		\includegraphics[width=\columnwidth]{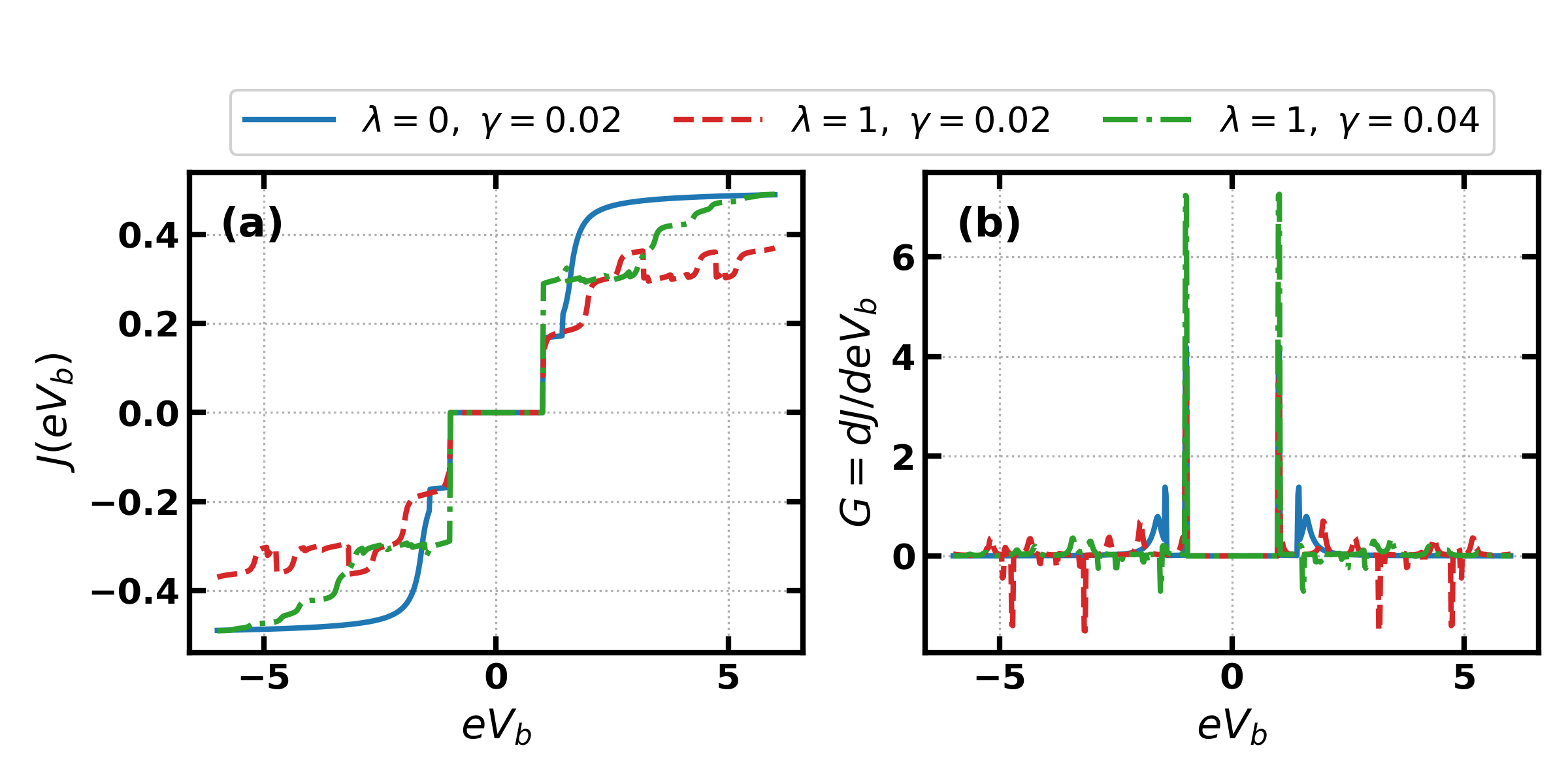}
		\label{fig:Current_Conductance_U0}
	}
	
	\vspace{0.4cm}
	
	\subfigure[Current $J(eV_b)$ and differential conductance $G = dJ/d(eV_b)$ for $U \neq 0$.]{
		\includegraphics[width=\columnwidth]{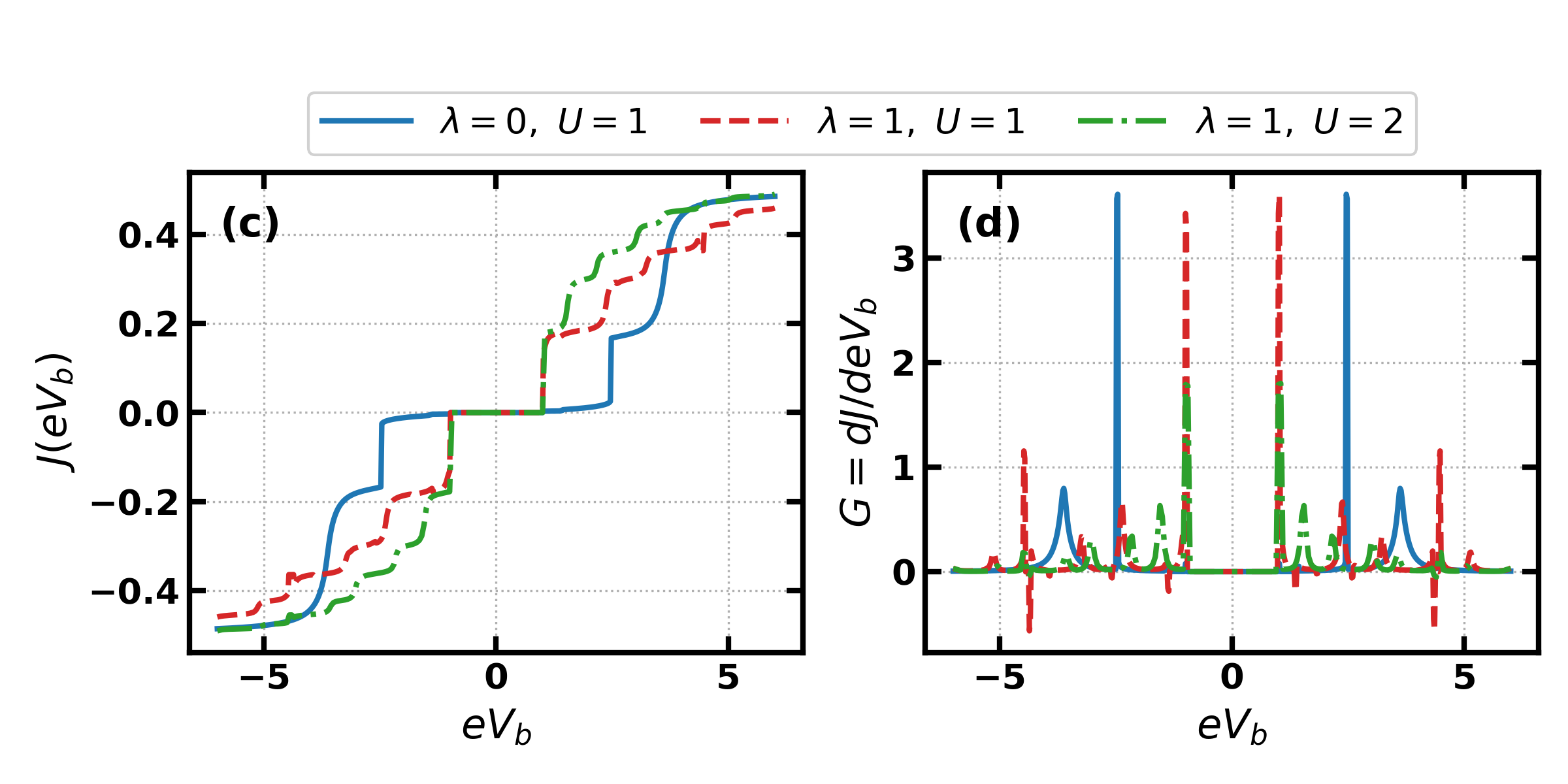}
		\label{fig:Current_Conductance_Ufinite}
	}
	
	\caption{
		Current $J(eV_b)$ and differential conductance $G = dJ/d(eV_b)$ as functions of the bias voltage $eV_b$.
		(a) Results for $U = 0$.  
		(b) Results for $U \neq 0$, showing the influence of electron--electron and electron--phonon interactions.
	}
	\label{fig:Current_Conductance_combined}
\end{figure}

	We have assumed that QD$_{1,2}$ is symmetrically connected to both the source and the drain. 
The energy level of each quantum dot (QD) is taken to be zero, and all energies are measured in units of the phonon energy $\hbar \omega_0$. 
For most of our calculations, we consider $\Gamma = 0.2$, $eV_g = 0$, $eV_m = 0.5$, $eV_b = 0.5$,$t_1=t_2 = t_3 = 0.3$ and $U = 1$. 
The Coulomb interaction is treated within the mean-field Hartree--Fock approximation. 
Because of the electron--phonon (e--p) interaction, the onsite Coulomb interaction energy $U$ is substantially reduced due to the polaronic effect, 
and thus the use of the mean-field approximation is justified for moderately large Coulomb interactions. 
Hence, our results lie well outside the Kondo regime. 
We also assume that the electron charge density in both the source and the drain remains constant.

Fig~\ref{fig:SpectralFunction} displays the behaviour of the spectral function $A(E)$ as a function of energy $E$ for different values of electron--phonon coupling strength $\lambda$ and dissipation parameter $\gamma$.  $\lambda = 0, \gamma = 0$ (blue solid line) corresponds to the non-interacting case, which shows two sharp peaks representing the discrete electronic levels of the isolated system. 
When $\lambda = 1$ and $\gamma = 0$ (orange dashed line), phonon sidebands appear, resulting from electron--phonon interactions that split the spectral weight into multiple peaks. 
As  $\gamma$ (green dot-dashed and red dotted lines) becomes finite, dissipation occurs because of substrate QDs interactions, which leads to broadening and reduction in intensity of these peaks. Thus, $\lambda$ creates multiple phonon-assisted channels, while $\gamma$ induces relaxation and dephasing, both reshaping the molecular density of states.

\begin{figure}[t]
	\centering
	\includegraphics[width=0.95\columnwidth]{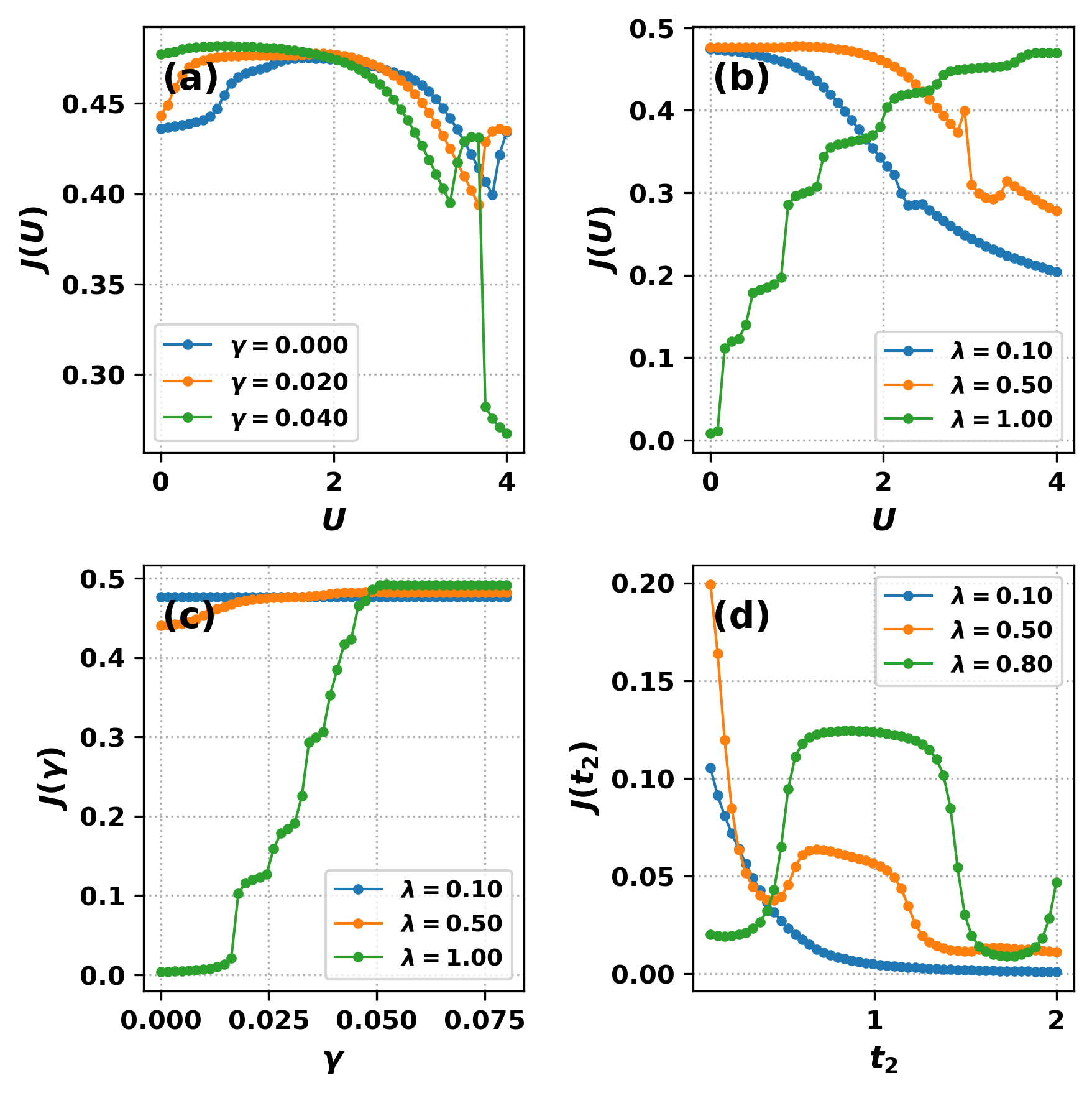}
	\caption{(a–d) Current $J$ as a function of interaction parameters. 
		(a) Variation of $J$ with on-site Coulomb interaction $U$ for different dissipation strengths $\gamma$. 
		(b) $J(U)$ for different electron–phonon couplings $\lambda$ at fixed $\gamma$. 
		(c) Dependence of $J$ on the phonon damping parameter $\gamma$ for several $\lambda$ values. 
		(d) Dependence of $J$ on inter-dot tunneling $t_2$ for different $\lambda$. 
		The data illustrate the interplay between Coulomb interaction, electron–phonon coupling, and dissipative effects in determining the transport characteristics of the TMT device.}
	\label{fig:J_vs_parameters}
\end{figure}

In Fig.~\ref{fig:J_vs_parameters}, we present the steady-state current $J$ as a function of different interaction parameters. 
Panel~(a) shows the influence of the dissipation strength $\gamma$, where increasing $\gamma$ suppresses the current owing to enhanced energy relaxation into the phonon bath. 
In panel~(b), we vary the electron–phonon coupling $\lambda$ at fixed $\gamma$. 
As $\lambda$  becomes larger, the current drops significantly and can even change in a non-smooth way. This happens because the electron–phonon interaction (polaron effect) weakens the effective tunneling between the quantum dots and the leads.
Panel~(c) displays the variation of $J$ with the phonon damping parameter $\gamma$, showing that for small $\gamma$, dissipation assists in stabilizing transport, but at larger $\gamma$ the current saturates due to excessive dephasing and phonon-induced localization effects. 
In panel~(d), the dependence on inter-dot tunneling $t_2$ is shown for various $\lambda$ values. 
At weak $\lambda$, the current rises smoothly with $t_2$, indicating coherent inter-dot transport. 
However, when $\lambda$ is large, oscillatory features and suppression appear, signaling phonon-mediated blockade and incoherent tunneling. 

Notably, panels~(b) and (d) exhibit regions of \emph{negative differential resistance} (NDR),
where the differential conductance $G = dJ/dV_b$ becomes negative.
In the trimer molecular transistor (TMT), this NDR arises from the interplay between
quantum interference within the triangular geometry,
Coulomb blockade effects, and strong electron–phonon interactions.
At higher bias, the redistribution of spectral weight among the molecular orbitals
and phonon sidebands effectively suppresses the transmission through certain paths,
leading to a reduction of current with increasing bias voltage.
Such phenomena are consistent with experimental and theoretical observations in
multi-dot and molecular devices~\cite{Kubatkin2003, Donarini2009, Sanchez2010, Jiang2013, Galperin2005, Koch2005, Hettler2002, Datta1995}.

\begin{figure}[t]
	\centering
	\includegraphics[width=0.95\columnwidth]{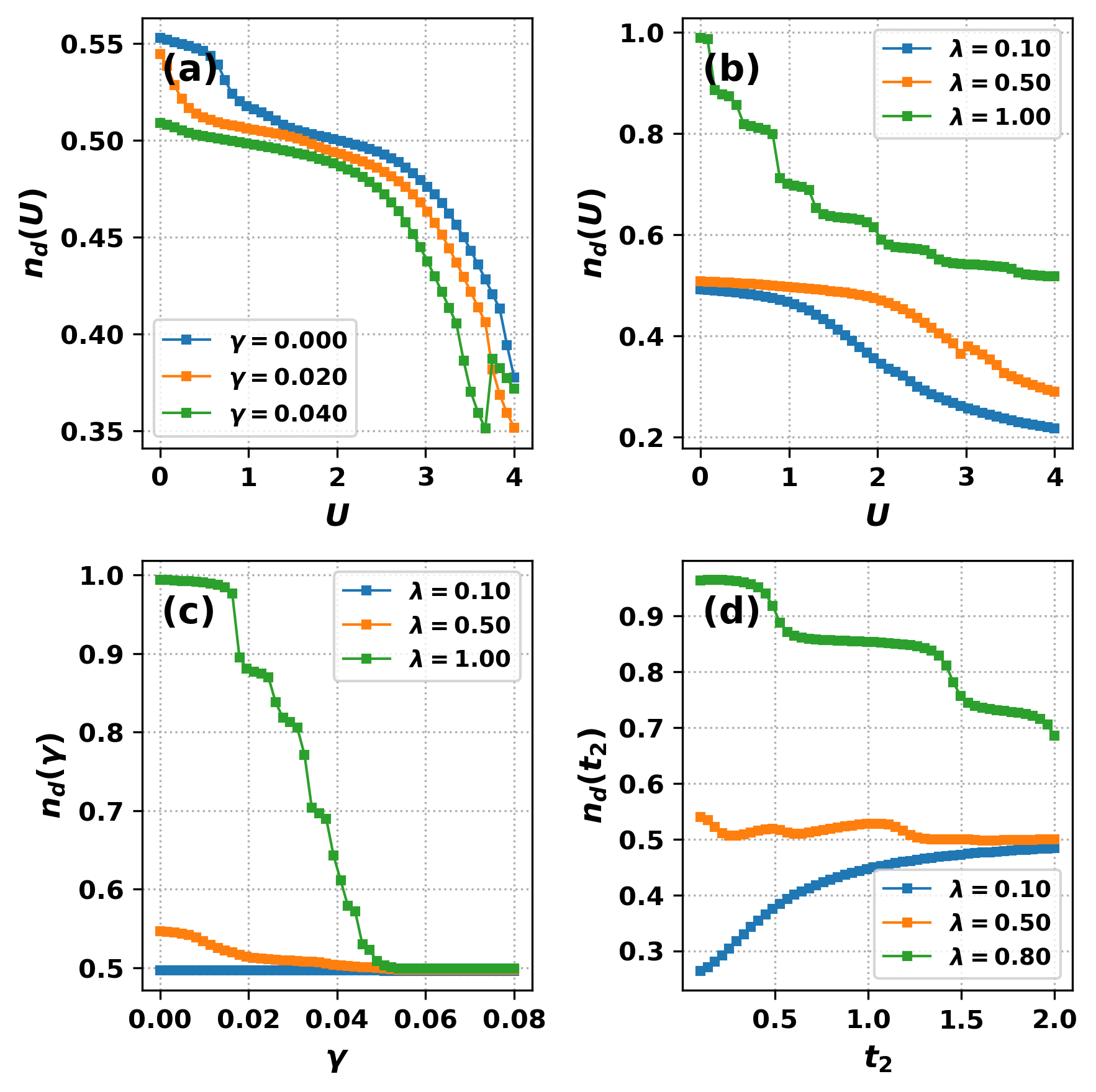}
	\caption{(a–d) Self-consistent dot occupation $n_d$ as a function of interaction parameters in the trimer molecular transistor (TMT). 
		(a) Dependence of $n_d$ on on-site Coulomb interaction $U$ for different dissipation strengths $\gamma$. 
		(b) $n_d(U)$ for several electron–phonon couplings $\lambda$ at fixed $\gamma$. 
		(c) Variation of $n_d$ with the phonon damping parameter $\gamma$ for different $\lambda$ values. 
		(d) Dependence of $n_d$ on inter-dot tunneling amplitude $t_2$ for various $\lambda$. 
		The results illustrate how Coulomb repulsion, electron–phonon coupling, and dissipation collectively influence the steady-state charge distribution within the TMT device.}
	\label{fig:ndu_vs_parameters}
\end{figure}

To further understand the microscopic origin of the current characteristics,
we analyze the self-consistent dot occupation $n_d = \langle d^\dagger d \rangle$,
which represents the steady-state electronic population on the molecular site.
Within the nonequilibrium Green’s function framework,
$n_d$ is obtained from the lesser Green’s function as
\[
n_d = \int \frac{A_{11}(E)f(E)}{2\pi}\, dE,
\]
where $A_{11}(E)$ is the local spectral function and $f(E)$ is the Fermi distribution.
The quantity $n_d$ measures the degree of charge filling on the dot:
$n_d \!\approx\! 0$ corresponds to an empty level,
$n_d \!\approx\! 1$ indicates full occupancy,
and intermediate values signify partial filling due to hybridization with the leads.

In Fig.~\ref{fig:ndu_vs_parameters}, we present the variation of $n_d$ as a function of key interaction parameters.
Panel~(a) shows $n_d(U)$ for several dissipation strengths $\gamma$. 
With increasing $U$, the occupation decreases monotonically,
reflecting the onset of Coulomb blockade that suppresses double occupancy.
For larger $\gamma$, this suppression becomes slightly stronger as the bath
facilitates energy relaxation and enhances localization.
Panel~(b) shows $n_d(U)$ for different electron–phonon couplings $\lambda$ at fixed $\gamma$. 
At small $\lambda$, the dot remains moderately occupied,
while larger $\lambda$ increases $n_d$ owing to polaronic self-trapping of electrons.
However, at high $U$, Coulomb repulsion again dominates,
leading to a decline in occupation.
Panel~(c) displays the dependence of $n_d$ on the phonon damping parameter $\gamma$.
For weak $\lambda$, $n_d$ remains almost constant,
but for stronger $\lambda$, a sharp decrease appears beyond a critical $\gamma_c$,
signaling a crossover from a coherent polaronic state to a dissipative regime dominated by phonon relaxation.
Finally, panel~(d) shows the dependence of $n_d$ on the inter-dot tunneling amplitude $t_2$.
At small $\lambda$, $n_d$ grows smoothly with $t_2$, consistent with coherent charge delocalization.
For larger $\lambda$, however, oscillatory and suppressed behavior emerges,
reflecting phonon-mediated blockade and incoherent tunneling.

Overall, the behavior of $n_d$ complements the current characteristics shown in Fig.~\ref{fig:J_vs_parameters}.
While the current reflects charge transport across the device,
$n_d$ captures the underlying charge accumulation within the trimer region.
Their combined trends reveal the delicate interplay between Coulomb repulsion,
electron–phonon coupling, and phonon damping.
Polaron formation enhances local occupation but suppresses current,
whereas dissipation and strong $U$ reduce both, driving the system toward
a localized and weakly conducting state.
These results highlight how vibrational coupling and dissipative effects
jointly shape charge transport and localization in nanoscale molecular transistors,
in agreement with earlier studies on polaronic and dissipative quantum transport~\cite{Lang1963, Caldeira1983, Meir1992, Koch2005, Galperin2007, Weiss2012, Sharma2024}.
\begin{figure}[t]
	\centering
	\includegraphics[width=0.95\columnwidth]{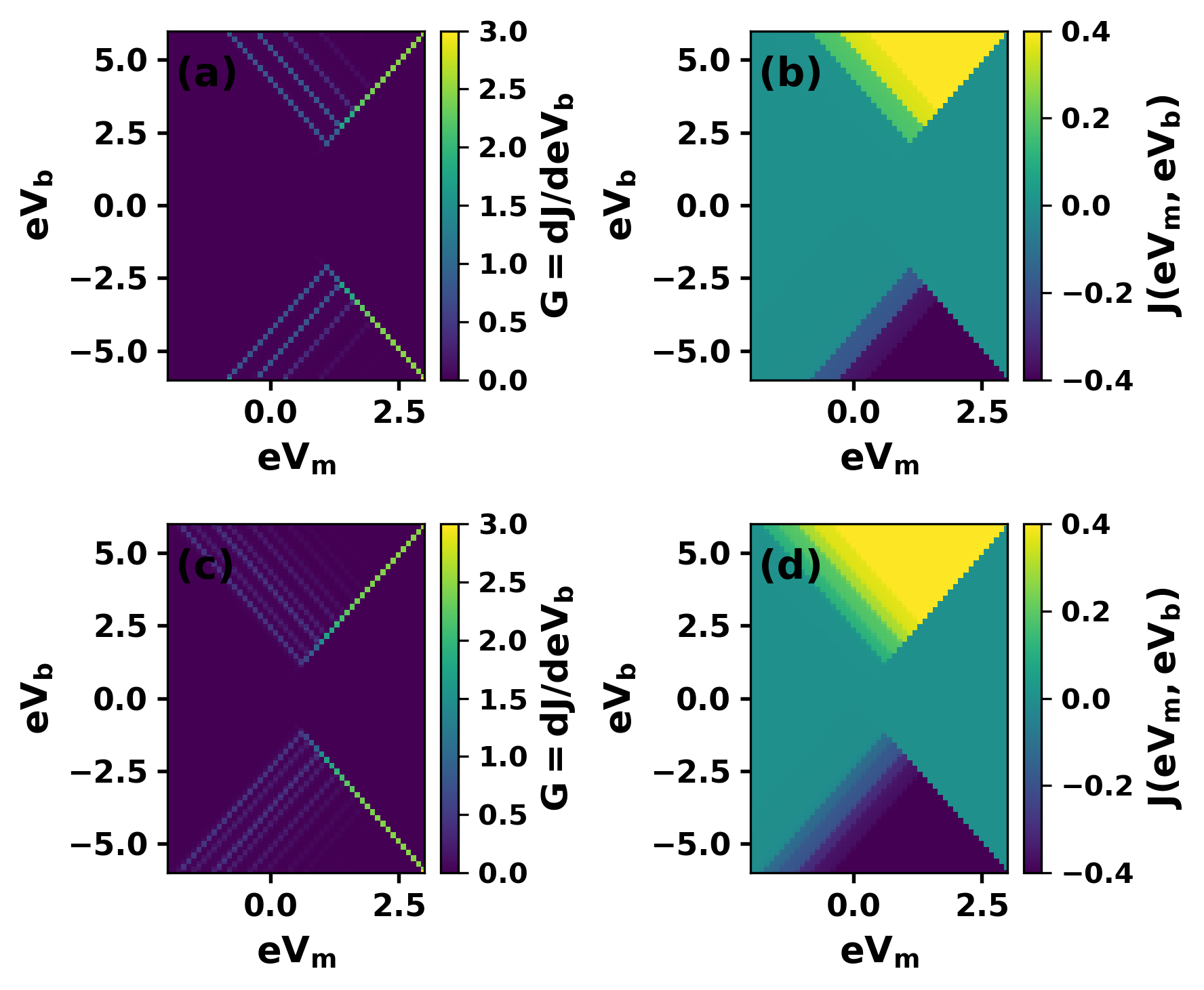}
	\caption{Color maps of (a,c) differential conductance $G = dJ/d(eV_b)$ 
		and (b,d) current $J(eV_m, eV_b)$ as functions of the mean and bias voltages $(eV_m, eV_b)$
		for the trimer molecular transistor (TMT). 
		Panels~(a,b) correspond to the noninteracting case ($U = 0$),
		while panels~(c,d) include a finite Coulomb interaction ($U \neq 0$).
		The results illustrate the evolution from purely resonant transport
		to Coulomb-blockade–split sidebands and nonlinear current response.}
	\label{fig:G_J_maps}
\end{figure}
Figure~\ref{fig:G_J_maps} displays the two-dimensional maps of 
differential conductance $G = dJ/d(eV_b)$ [panels~(a,c)] 
and steady-state current $J(eV_m, eV_b)$ [panels~(b,d)] 
as functions of the mean voltage $eV_m$ and bias voltage $eV_b$.
The upper panels~(a,b) correspond to the noninteracting case ($U = 0$),
whereas the lower panels~(c,d) represent results for finite on-site Coulomb interaction ($U \neq 0$).
These plots reveal how electron correlations modify the transport window and current pathways in the trimer molecular transistor.

In the noninteracting case [Figs.~\ref{fig:G_J_maps}(a,b)],
transport occurs through a pair of resonant channels aligned with the molecular levels of the trimer.
The conductance map in panel~(a) shows sharp diagonal ridges at 
$eV_b = \pm 2(eV_m - \varepsilon_d)$,
which correspond to the onset of resonant tunneling as the chemical potentials of the leads align with the molecular states.
The bright lines reflect enhanced transmission and are symmetric about $eV_b = 0$,
characteristic of coherent transport through delocalized molecular orbitals.
The corresponding current map in panel~(b) exhibits triangular high-current regions bounded by the same resonance lines,
indicating that current increases linearly within the transport window and saturates once both levels lie outside the bias window.

When Coulomb interaction is introduced [Figs.~\ref{fig:G_J_maps}(c,d)],
the spectra split due to charging effects, and the transport pattern changes dramatically.
In the conductance map of panel~(c),
additional diagonal ridges appear, separated by approximately $U$,
reflecting the emergence of Coulomb-split sidebands.
These new lines correspond to transitions between singly and doubly occupied states,
which are shifted in energy by the on-site repulsion.
Consequently, the conductance peaks become weaker and more widely spaced,
indicating suppression of resonant tunneling and the onset of Coulomb blockade.
The current map in panel~(d) shows similar features:
the bright triangular regions shrink and move to higher $|eV_b|$,
while the slope of $J(eV_b)$ becomes less steep due to reduced tunneling probability.
This behavior demonstrates that electron–electron interactions not only reduce the overall current
but also shift the transport thresholds to higher bias.

Overall, the transition from the upper to lower panels highlights the evolution from coherent,
resonant transport in the noninteracting regime to correlation-dominated transport in the interacting case.
The formation of Coulomb gaps, visible as dark regions between the bright ridges,
marks the Coulomb-blockade regime, where current is suppressed until sufficient bias is applied to overcome $U$.
Such patterns are hallmarks of correlated electron transport in quantum-dot arrays and molecular devices~\cite{Meir1992, Hanson2007, Trocha2010, Koch2005, Galperin2007}.

\section{Summary}

In summary, we have studied how charge flows through a trimer molecular transistor, taking into account electron–phonon interactions, phonon damping, and on-site Coulomb repulsion (treated using a mean-field approach).
Using the Lang–Firsov transformation combined with the nonequilibrium Green’s function formalism, 
we captured polaronic renormalization, phonon sidebands, and dissipative broadening within a unified framework.
Our analysis reveals that:
(i) electron–phonon coupling $\lambda$ induces phonon-assisted sidebands and eventual self-trapping;
(ii) dissipation $\gamma$ suppresses coherence and stabilizes current at weak coupling but leads to localization at strong damping;
and (iii) Coulomb interaction $U$ produces charging-induced suppression and Coulomb-split resonances.
The interplay of these mechanisms gives rise to negative differential resistance, spectral redistribution, and correlation-driven current suppression.
The resulting maps of $J(eV_m,eV_b)$ and $G(eV_m,eV_b)$ distinctly display the transition from coherent resonant tunneling to the Coulomb-blockade regime.
These results give a detailed microscopic understanding of how phonon coupling, dissipation, and electron–electron interactions work together to control charge transport in nanoscale molecular transistors and quantum-dot arrays. They also suggest possible ways to tune and optimize conductance in future molecular electronic devices.

	\bibliographystyle{apsrev4-2}

\end{document}